%% file: article.tex
\documentclass[twocolumn,twoside]{IEEEtran}
\usepackage{mathpazo}
\usepackage{times}
\usepackage{amsmath}
\usepackage{amsfonts}
\usepackage{latexsym}
\usepackage{amssymb}
\usepackage{cite}

\usepackage{upref}
\usepackage{theorem}
\usepackage{graphicx}
\usepackage{psfrag}

\usepackage[usenames]{color}
\usepackage{overpic}

\theoremstyle{plain}
\theorembodyfont{\normalfont\slshape}

\newtheorem{thm}{Theorem$\!$}
\newenvironment{theorem}
{\begin{thm}\hspace*{-1ex}{\bf.}}{\end{thm}}

\newtheorem{lem}[thm]{Lemma$\!$}
\newenvironment{lemma}{\begin{lem}\hspace*{-1ex}{\bf.}}{\end{lem}}

\newtheorem{prop}[thm]{Proposition$\!$}

\newtheorem{cor}[thm]{Corollary$\!$}
\newenvironment{corollary}{\begin{cor}\hspace*{-1ex}{\bf.}}{\end{cor}}

\newtheorem{defn}[thm]{Definition$\!$}
\newenvironment{definition}{\begin{defn}\hspace*{-1ex}{\bf.}}{\end{defn}}

\newtheorem{xmpl}[thm]{Example$\!$}
\newenvironment{example}{\begin{xmpl}\hspace*{-1ex}{\bf.}}{\hfill$\Box$\end{xmpl}}

\newtheorem{cnstr}{Construction$\!$}

\newenvironment{construction}{\begin{cnstr}\hspace*{-1ex}{\bf.}}{\hfill$\Box$\end{cnstr}}

\newcounter{enumrom}
\renewcommand{\theenumrom}{(\roman{enumrom})}

\makeatletter
\renewcommand{\@endtheorem}{\endtrivlist}
\makeatother

\makeatletter
\renewcommand{\thefigure}{{\@arabic\c@figure}}
\renewcommand{\fnum@figure}{{\bf Figure\,\thefigure}}
\makeatother

\newcommand{\cC}{\mathcal{C}}

\newcommand{\cE}{\mathcal{E}}
\newcommand{\cF}{\mathcal{F}}
\newcommand{\cG}{\mathcal{G}}
\newcommand{\cH}{\mathcal{H}}

\newcommand{\cK}{\mathcal{K}}
\newcommand{\cL}{\mathcal{L}}
\newcommand{\cM}{\mathcal{M}}
\newcommand{\cN}{\mathcal{N}}

\newcommand{\cP}{\mathcal{P}}

\newcommand{\cT}{\mathcal{T}}

\newcommand{\cV}{\mathcal{V}}

\newcommand{\mathset}[1]{\left\{#1\right\}}
\newcommand{\abs}[1]{\left|#1\right|}

\newcommand{\parenv}[1]{\left( #1 \right)}

\newcommand{\be}[1]{\begin{equation}\label{#1}}
\newcommand{\ee}{\end{equation}}

\renewcommand{\leq}{\leqslant}

\renewcommand{\geq}{\geqslant}

\renewcommand{\Bbb}{\mathbb}

\newcommand{\Cref}[1]{Co\-ro\-lla\-ry\,\ref{#1}}

\renewcommand{\Bbb}{\mathbb}

\newcommand{\F}{{\Bbb F}}

\newcommand{\N}{\mathbb{N}}
\newcommand{\R}{\mathbb{R}}

\newcommand{\bP}{\mathbb{P}}

\newcommand{\eqdef}{\triangleq}

\DeclareMathOperator{\inc}{In}
\DeclareMathOperator{\rank}{rank}
\DeclareMathOperator{\skel}{skel}
\DeclareMathOperator{\gap}{gap}
\newcommand{\din}{d_{\mathrm{in}}}
\newcommand{\sbinom}[2]{\genfrac{[}{]}{0pt}{}{#1}{#2}}
\newcommand{\oq}{\overline{q}}
\newcommand{\ot}{\overline{t}}

\outer\def\proclaim #1. #2\par{\medbreak
 \noindent{\bf#1.\enspace}{\sl#2\par}%
 \ifdim\lastskip<\medskipamount \removelastskip\penalty55\medskip\fi}

\usepackage{pgf}
\usepackage{tikz}
\usetikzlibrary{matrix,chains,positioning,arrows,calc,decorations,plotmarks,patterns, fit,backgrounds}
\usepackage{pgfplots}
\usetikzlibrary{external}                       
\usepackage{tkz-graph}
\pgfplotsset{compat=1.3}
\tikzstyle{help lines}=[black!20,dashed]

\pgfplotscreateplotcyclelist{mylist}{red,blue,black,yellow,brown}

\pgfdeclarepatternformonly{my north east lines}{\pgfqpoint{-1pt}{-1pt}}{\pgfqpoint{8pt}{8pt}}{\pgfqpoint{6pt}{6pt}}%
{
	\pgfsetlinewidth{0.4pt}
	\pgfpathmoveto{\pgfqpoint{0pt}{0pt}}
	\pgfpathlineto{\pgfqpoint{6.1pt}{6.1pt}}
	\pgfusepath{stroke}
}
\definecolor{light_gray}{rgb}{0.6,0.6,0.6}
\definecolor{awgray}{rgb}{0.7,0.7,0.7}
\definecolor{awgray_dark}{rgb} {0.4,0.4,0.4}

\tikzset{
	>=stealth',
	mycircle/.style={circle, draw=gray, very thick, text width=.1em, minimum height=1.5em, text centered},
	mycircle_small/.style={circle,draw=awgray_dark,fill = awgray_dark, inner sep=0,minimum size=.6em},
	mycircle_small_black/.style={circle,draw=black,fill = black, inner sep=0,minimum size=.6em},
	mybox/.style={rectangle,rounded corners,draw=black, thick,text width=1em,minimum height=4em,minimum width=4em,text centered},
	mybox_small/.style={rectangle,rounded corners,draw=black, thick,text width=1em,minimum height=2em,minimum width=2em,text centered},
	mybox_vec/.style={rectangle,rounded corners,draw=black, thick,text width=1em,minimum height=0.7em, minimum width=4em,text centered},
	mybox_vec_short/.style={rectangle,rounded corners,draw=black, thick,text width=1em,minimum height=0.7em, minimum width=2em,text centered},
	pil/.style={->, thick, shorten <=2pt, shorten >=2pt,},
}

\begin{document}


\title{\textbf{Network-Coding Solutions for Minimal Combination Networks and Their Sub-networks}}


\author{
  Han Cai,~\IEEEmembership{Member,~IEEE},
  Johan Chrisnata,
  Tuvi Etzion,~\IEEEmembership{Fellow,~IEEE},\\
  Moshe Schwartz,~\IEEEmembership{Senior Member,~IEEE},
  and
  Antonia Wachter-Zeh,~\IEEEmembership{Member,~IEEE}
  \thanks{The material in this paper was presented in part at the International Symposium on Information Theory (ISIT), Paris, France, July 2019.}%
  \thanks{Han Cai is with the School
    of Electrical and Computer Engineering, Ben-Gurion University of the Negev,
    Beer Sheva 8410501, Israel
    (e-mail: hancai@aliyun.com).}%
  \thanks{Johan Chrisnata is with the School of Physical and Mathematical Sciences, Nanyang Technological University, Singapore, and the Department of Computer Science, Technion -- Israel Institute of Technology, Haifa 3200003, Israel, (e-mail: johan.c@cs.technion.ac.il).}%
  \thanks{Tuvi Etzion is with the Department of Computer Science, Technion -- Israel Institute of Technology, Haifa 3200003, Israel, (e-mail: etzion@cs.technion.ac.il).}%
  \thanks{Moshe Schwartz is with the School
    of Electrical and Computer Engineering, Ben-Gurion University of the Negev,
    Beer Sheva 8410501, Israel
    (e-mail: schwartz@ee.bgu.ac.il).}%
  \thanks{Antonia Wachter-Zeh is with the
    Institute for Communications Engineering, Technical University of Munich,
    80333 Munich, Germany
    (e-mail: antonia.wachter-zeh@tum.de).}%
  \thanks{A.~Wachter-Zeh and M.~Schwartz were supported in part by
a German Israeli Project Cooperation (DIP) grant under grant no.~PE2398/1-1 and KR3517/9-1.}
}

\maketitle

\begin{abstract}
Minimal multicast networks are fascinating and efficient combinatorial
objects, where the removal of a single link makes it impossible for
all receivers to obtain all messages. We study the structure of such
networks, and prove some constraints on their possible solutions.
  
We then focus on the combination network, which is one of the simplest
and most insightful network in network-coding theory. Of particular
interest are minimal combination networks. We study the gap in
alphabet size between vector-linear and scalar-linear network-coding
solutions for such minimal combination networks and some of their
sub-networks.

For minimal multicast networks with two source messages we find the
maximum possible gap. We define and study sub-networks of the
combination network, which we call Kneser networks, and prove that
they attain the upper bound on the gap with equality. We also prove
that the study of this gap may be limited to the study of sub-networks
of minimal combination networks, by using graph homomorphisms
connected with the $q$-analog of Kneser graphs. Additionally, we prove
a gap for minimal multicast networks with three or more source
messages by studying Kneser networks.

Finally, an upper bound on the gap for full minimal combination networks
shows nearly no gap, or none in some cases. This is obtained using an
MDS-like bound for subspaces over a finite field.
\end{abstract}



\section{Introduction}
\label{sec:intro}

\IEEEPARstart{N}{etwork} coding has been attracting increased
attention for almost two decades since the seminal
papers~\cite{AhlCaiLiYeu00,LiYeuCai03}. Multicast networks have
received most of this attention.  A recent survey on the foundations
of multicast network coding may be found in~\cite{FraSol16}.  The
multicast network-coding problem can be formulated as follows: given a
network with one source which has $h$ messages, for each edge find a
function of the messages received at the starting node of the edge,
such that each terminal can recover all the messages from its received
messages.  Such an assignment of a function to each edge is called a
\emph{solution} for the network.

Obviously, received messages on an edge can be expressed as functions
of the source messages.  If these functions are linear, we obtain a
\emph{linear solution}. Otherwise, we have a \emph{nonlinear
  solution}.  In linear network coding, each linear function on an
edge consists of coding coefficients for each incoming message.  If
the coding coefficients and the messages are scalars, it is called a
\emph{scalar solution}.  If the messages are vectors
and the coding coefficients are matrices then it is called a
\emph{vector solution}.  A network which has a solution
is called a \emph{solvable} network.  It is well-known that a
multicast network with one source, $h$ messages, and $N$ terminals, is
solvable if and only if the min-cut between the source and each
terminal is at least $h$~\cite{FraSol16}.

The minimal \emph{alphabet size}, and in the linear setting,
\emph{field size}, of a solution is an important parameter that
directly influences the complexity of the calculations at the network
nodes. An efficient algorithm to find a field size (not necessarily
minimal) that allows a linear solution, and the related linear network
code was given in~\cite{JagSanChoEffEgnJaiTol05}. It is known that any
field size $q\geq N$ suffices for a linear solution, but it is
conjectured that the smallest field size allowing a linear solution is
much smaller \cite{FraSol06,FraSol16}. This, however, was proved only
for two messages~\cite{FraSol06}.

In this work we distinguish between scalar and vector linear
solutions. Given a network $\cN$, we define $q_s(\cN)$ to be the
smallest field size $q$ for which $\cN$ has a scalar linear
solution. Similarly, we define $q_v(\cN)$ to be the smallest value
$q^t$, $q$ a prime power, such that $\cN$ has a vector solution with
vectors of length $t$ over a field of size $q$. By definition,
\[q_s(\cN)\geq q_v(\cN),\]
and we define the \emph{gap} by
\[
\gap (\cN)\eqdef q_s(\cN)-q_v(\cN).
\]

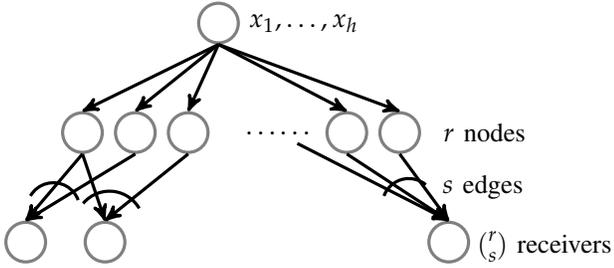
\begin{figure}[tb]
  \input{comb-network.tex}
  \caption{The $\mathcal{N}_{h,r,s}$ (full) combination network: it
    has an edge from the source to each of the $r$ nodes in the middle
    layer. Each of the $\binom{r}{s}$ terminals is connected to a
    unique set of $s$ middle-layer nodes, and wants to recover all of
    the $h$ source messages.  }
  \label{fig:comb-net}
\end{figure}

One of the most celebrated families of networks is the family of
combination networks~\cite{RiiAhl06}, which has been used for various
topics in network coding, e.g.,
\cite{GheBidFraTol11,XiaMedAul07,NgaYeu04,MahLiLi12,GuoShiCaiMed13}.
The $\mathcal{N}_{h,r,s}$ combination network, where $s \geq h$, is
shown in Fig.~\ref{fig:comb-net}.  The network has three layers: the
first layer consists of a single source with $h$ messages.  The source
transmits $r$ messages to the $r$ nodes of the middle layer. Any $s$
nodes in the middle layer are connected to a terminal, and each one of
the $\binom{r}{s}$ terminals wants to recover all the $h$
messages. Since we shall also examine sub-networks of the combination
network, we will sometimes stress that no part of the network has been
removed by saying that the network is a \emph{full} combination
network.

It was proved in~\cite{RiiAhl06} that a solution for a combination
network exists if and only if a related error-correcting code
exists. This network was also generalized to compare scalar and vector
network coding~\cite{EtzWac18}.  Its sub-networks were used to prove
that finding the minimum required field size of a (linear or
nonlinear) scalar network code for a certain multicast network is
NP-complete~\cite{LehLeh04}.

Of particular interest are \emph{minimal multicast networks}. In such
a network, the removal of even a single link reduces the cut between
the source and at least one of the terminals, which in turn, makes the
network unable to transmit all of its messages to all the terminals,
i.e., the network becomes unsolvable. In the case of combination
networks, only the $\cN_{h,r,h}$ networks are minimal. Minimal
networks are not only a fascinating extremal combinatorial objects,
but also of practical importance since they require the least amount
of network resources to enable the multicast operation. Minimal
networks have been studied in the past, e.g.,
\cite{LehLeh04,ElRGeoSpr06}, but only in the context of scalar network
coding.

The goal of this work is to consider two problems which are related to
vector coding solutions for minimal combination networks and their
sub-networks. Our main contributions are the following: we first study
general multicast networks which are minimal. We extend the scalar
setting of \cite{LehLeh04,ElRGeoSpr06} to the vector setting, and we
show minimality entails some structural properties of the graph, as
well as some constraints on linear solutions.

We then focus on the gap in general minimal multicast networks.  The
first to demonstrate a gap exists in some network was
\cite{SunYanLonYinLi16}. However, the gap there is only $1$, and
requires at least $h\geq 8190$ source messages.  This was
significantly improved by \cite{EtzWac18}, which showed much larger
gaps for as little as $h\geq 4$ source messages, in non-minimal
networks. In this paper we show that large gaps exist already for
$h=2$ messages, the lowest number of messages for which a gap is
possible. Not only that, but the networks we construct are minimal,
and we show that they attain the highest possible gap of all minimal
multicast networks with two source messages. We further prove that
studying the gap in general minimal multicast networks with two source
messages is equivalent to studying it only in minimal multicast
networks which are sub-networks of combination networks, thus, further
motivating the study of combination networks. The main tool we use is
a connection with $q$-Kneser graphs (which are $q$-analogs of Kneser
graphs), as well as a generalization to a new $q$-Kneser hypergraph.

Finally, we also prove an upper bound on the gap in full minimal
combination networks. To the best of our knowledge, this is the first
such upper bound on the gap. The bound is obtained using certain
longest MDS array codes (e.g., see \cite{BlaBruVar96}) or a
combinatorial structure which we call an independent
configuration. While an upper bound on the length of such codes
already exists, we present a different proof approach based on the
properties of subspaces in the configuration.

The paper is organized as follows. In Section~\ref{sec:prem} we
provide the basic notation and definitions used in the
paper. Section~\ref{sec:minimal} studies fundamental properties of
general minimal multicast networks. In Section~\ref{sec:twomessages}
we study the gap in networks with two source messages, whereas
Section~\ref{sec:moremessages} is devoted to three or more source
messages.  Section~\ref{sec:mincomb} focuses on an upper bound on the
gap for full minimal combination networks. We conclude in
Section~\ref{sec:conclude} with a summary of the results and some open
problems.

\section{Preliminaries}
\label{sec:prem}

We now provide the basic notation and definitions used throughout the
paper. If $W$ is some finite set of $n$ elements, we use
$\binom{W}{t}$ to denote the set of all subsets of $W$ of size
$t$. Obviously, using the binomial coefficients,
\[\abs{\binom{W}{t}}=\binom{n}{t}\eqdef\frac{n!}{t!(n-t)!}.\]

Now, let $\F_q$ denote the finite field of size $q$, where $q\in\bP$
and $\bP\subset\N$ denotes the set of prime powers. Taking the
$q$-analog of sets, if $V$ is a vector space over $\F_q$, with
$n=\dim V$, we use $\sbinom{V}{t}$ to denote the set of all vector
subspaces of $V$ of dimension $t$. It is well known that the size of
$\sbinom{V}{t}$ is given by the Gaussian coefficient, namely,
\[\abs{\sbinom{V}{t}}=\sbinom{n}{t}_q\eqdef\frac{\prod_{i=1}^n(q^i-1)}{\prod_{i=1}^t(q^i-1)\prod_{i=1}^{n-t}(q^i-1)}.\]
We omit the subscript $q$ whenever the field size is understood from
the context, and just write $\sbinom{n}{t}$.

For any $x\in\R$, $x>0$ we use $\psi(x)$ to denote the smallest power of a
prime that is greater or equal to $x$, i.e.,
\begin{equation}
  \label{eq:psidef}
  \psi(x) \eqdef \min \mathset{ q\in\bP ~:~ q\geq x }.
\end{equation}
By Bertrand's postulate (e.g., see \cite{Apo76}),
\begin{equation}
  \label{eq:primelarge}
  0\leq \psi(n)-n \leq n,
\end{equation}
for all $n\in\N$. We mention that much stronger results may be
obtained at the cost of working only for large enough $n$. For
example, \cite{BakHarPin01} showed that the interval $[x,x+x^{21/40}]$
contains a prime for all large enough $x$. Thus, for all large enough
$n$,
\begin{equation}
  \label{eq:primesmall}
  0\leq \psi(n)-n\leq n^{21/40}.
\end{equation}

Our main objects of interest are networks. We shall always assume that
our network consists of a finite directed acyclic graph
$\cG=(\cV,\cE)$. Nodes in the graph will be denoted using lower-case
Greek letters. The network contains a single node that is designated
as the source, $\sigma\in\cV$. It also contains terminal nodes
$\cT=\mathset{\tau_1,\tau_2,\dots,\tau_N}\subseteq\cV$. Finally, the
source has $h$ messages, denoted $x_1,\dots,x_h\in\F_q^t$.  We
therefore denote this network by $\cN=(\cG,\sigma,\cT,h)$.

In the network-coding model, each edge $e\in\cE$ carries a value from
$\F_q^t$. For a node $\nu\in\cV$, let $\din(\nu)$ denote the
in-degree of $\nu$, and let $\inc(\nu)$ denote the set of incoming
edges into $\nu$, hence $\abs{\inc(\nu)}=\din(\nu)$. For each node
$\nu$ and each edge $e$ outgoing from $\nu$, the value $e$ carries is
a function of the values on edges incoming into $\nu$. In a
\emph{linear} setting, this value is a linear combination of the
incoming values, where the coefficients are $t\times t$ matrices over
$\F_q$.

The goal of the terminals is to gain knowledge of source messages. In
the \emph{multicast} setting, every terminal wants to recover all the
source messages. If each of the terminals may recover all the source
messages then a \emph{linear solution} to the network is possible, and
we say $\cN$ has a $(q,t)$-linear solution.  When $t=1$ we call it a
\emph{scalar linear solution}, and in general, a \emph{vector linear
  solution}.

The source messages naturally form a vector space over $\F_q$ of
dimension $ht$, namely, $\F_q^{ht}$. In the linear case, for each
edge $e\in\cE$, the values carried by $e$ also form a vector space
over $\F_q$, which we denote by $\cM(e)$. More precisely, this
vector space is given by
\[\cM(e)=\mathset{G_e\cdot(x_1 |  \dots | x_h)^T ~:~ x_i\in\F_q^t},\]
where $G_e$ is a $t\times ht$ matrix over $\F_q$. The matrix $G_e$
is called the \emph{global coding matrix} (and when $t=1$, the
\emph{global coding vector}). Obviously,
\[0\leq \dim \cM(e) =
\rank(G_e) \leq t.\]

We also say each vertex $\nu\in\cV$ has access to
\begin{equation}
  \cM(\nu)\eqdef \mathset{G_{\nu}\cdot(x_1 |  \dots | x_h)^T ~:~ x_i\in\F_q^t}, \label{eq:defM}
\end{equation}
where $G_{\nu}$ is the $(\din(\nu)\cdot t)\times ht$ matrix over
$\F_q$ defined by
\[ G_{\nu}\eqdef \begin{pmatrix} G_{e_1} \\ G_{e_2}\\ \vdots \\ G_{e_{\din(\nu)}}\end{pmatrix},\]
with $\inc(\nu)=\mathset{e_1,\dots,e_{\din(\nu)}}$.  Thus,
\begin{equation}
  \label{eq:dimmv}
  \dim\cM(\nu)=\rank(G_{\nu})\leq
  \min(\din(\nu)\cdot t, ht).
\end{equation}
Combining these facts together, a terminal $\tau\in\cT$ is successful
in the linear multicast setting if and only if
\[\dim\cM(\tau)=\rank(G_{\tau})=ht.\]

A node in the network is said to be \emph{essential} if it is on a
path from the source to some terminal. We assume throughout the paper
that all nodes are essential, since non-essential nodes may be removed
without affecting the solution.

A \emph{cut} in the network is defined by a partition $\cV=S\cup T$,
with the source $\sigma\in S$, and some terminal $\tau_i\in T$. We say
the size of the cut is $m$ if there are exactly $m$ edges crossing it
from $S$ to $T$. It is well known (see \cite{FraSol16}) that in a
multicast network there exists a $(q,t)$-linear solution if and only
if every cut has size at least $h$.

Given a network $\cN=(\cG,\sigma,\cT,h)$, we define $q_s(\cN)$ to be
the smallest field size $q$ for which $\cN$ has a $(q,1)$-scalar
linear solution. Such a solution is said to be
\emph{scalar-optimal}. We also define $q_v(\cN)$ to be the smallest
value $q^t$ such that $\cN$ has a $(q,t)$-linear solution for some
$t$, and such a solution is said to be \emph{vector-optimal}. By
definition, $q_s(\cN)\geq q_v(\cN)$, and so we define the
\emph{(vector) gap} by
\[\gap(\cN)\eqdef q_s(\cN)-q_v(\cN).\]

Finally, as mentioned in the previous section, a celebrated family of
networks that has been studied extensively is the family of
combination networks. We briefly recall its definition: for
$h,r,s\in\N$, $r\geq s$, the $\cN_{h,r,s}$ combination network
(depicted in Fig.~\ref{fig:comb-net}) is a multicast network with a
single source $\sigma$, connected to $r$ nodes in the middle layer,
which we denote $\cL=\mathset{\lambda_1,\dots,\lambda_r}$. We then
have $\binom{r}{s}$ terminal nodes, which we may think of as indexed
by $\binom{\cL}{s}$. Each of the terminals is connected to a unique
subset of $s$ nodes from the middle layer. These networks and their
sub-networks will be the focus of this work.

\section{Minimal Multicast Networks}
\label{sec:minimal}

A \emph{minimal} multicast network can deliver $h$ messages from the
source to each of the terminals while each of its proper sub-networks
(containing all the original terminals) can deliver at most $h-1$
messages to at least one of the terminals.  From a practical point of
view, considering such minimal networks is interesting as it minimizes
the used network resources. From a theoretical perspective, minimal
networks are a fascinating extremal combinatorial object. Minimal
networks have been considered in the past \cite{LehLeh04,ElRGeoSpr06},
however only in the scalar case. The results of this section may be
considered as a generalization of these works.

\begin{definition}
  A multicast network $\cN=(\cG,\sigma,\cT,h)$ is said to be \emph{minimal}
  if every edge crosses a cut of size $h$.
\end{definition}

Thus, in a minimal network, the removal of any edge from $\cE$ makes
at least one cut have size strictly less than $h$, and therefore
unsolvable.

Given a directed graph $\cG=(\cV,\cE)$, and $m\in\N$, we define the
$m$-parallelized version of it as $m\cG=(\cV,m\cE)$, where $m\cE$
denotes the multiset obtained from $\cE$ by having each element appear
$m$ times. In essence, in the $m$-parallelized graph we keep the same
nodes, but duplicate each edge $m$ times.

\begin{lemma}
  \label{lem:split}
  If $\cN=(\cG,\sigma,\cT,h)$ is a minimal multicast network with a
  $(q,t)$-linear solution, then $\cN'=(t\cG,\sigma,\cT,ht)$ is a
  minimal multicast network with a $(q,1)$-scalar linear solution.
\end{lemma}
\begin{IEEEproof}
  The claim is entirely trivial: each transmitted vector on an edge
  $\nu_1\xrightarrow{e} \nu_2$ in $\cN$ is broken up into its $t$
  components which are then transmitted separately over the
  corresponding $t$ parallel edges in $\cN'$.  These are clearly
  linear combinations (with scalar coefficients from $\F_q$) of the
  values entering the node $\nu_1$ in $\cN'$. Finally, the size of
  each cut in $\cN'$ is obviously $t$ times the size of the same cut
  in $\cN$, hence the minimality of $\cN'$.
\end{IEEEproof}

\begin{lemma}
  \label{lem:dimmv}
  Consider a minimal multicast network $\cN=(\cG,\sigma,\cT,h)$ with a
  $(q,t)$-linear solution. Then
  \[\dim \cM(\nu)=\din(\nu)\cdot t\]
  for each $\nu\in\cV\setminus\mathset{\sigma}$, where $\cM(\nu)$ was
  defined in \eqref{eq:defM}.
\end{lemma}
\begin{IEEEproof}
  Define $\cN'=(t\cG,\sigma,\cT,ht)$, and construct a $(q,1)$-scalar
  linear solution to $\cN'$ from the solution to $\cN$, as described
  in the proof of Lemma~\ref{lem:split}. It is obvious that $\cM(\nu)$
  remains unchanged for all $\nu\in \cV$. To avoid confusion, we let
  $\din^{\cG}$ denote the in-degree in the graph $\cG$, and
  $\din^{t\cG}$ the in-degree in the $t$-parallelized graph
  $t\cG$. Assume to the contrary that $\dim
  \cM(\nu)<t\cdot\din^{\cG}(\nu)=\din^{t\cG}(\nu)$. Then, in the network
  $\cN'$ there exists an edge $e\in\inc(\nu)$ that always carries some
  fixed linear combination of other edges in $\inc(\nu)$. Removing $e$
  still allows a solution to $\cN'$ since the original $\cM(\nu)$ may
  still be computed from the surviving edges, and therefore the
  original values on edges leaving $\nu$ may be computed as a linear
  combination as well. However, this contradicts the minimality of
  $\cN'$ obtained by Lemma~\ref{lem:split}.
\end{IEEEproof}

The following corollary is trivial in the scalar regime, appearing as
a side note in \cite{ElRGeoSpr06}. Using the previous lemmas it may
also be proved for the general case.

\begin{corollary}
  \label{cor:din}
  In a minimal multicast network $\cN=(\cG,\sigma,\cT,h)$ with a
  $(q,t)$-linear solution we have $\din(\nu)\leq h$ for all
  $\nu\in\cV$.
\end{corollary}
\begin{IEEEproof}
  Combining Lemma~\ref{lem:dimmv} with \eqref{eq:dimmv} we get
  \[ \din(\nu)\cdot t = \dim \cM(\nu) \leq ht.\]
  The claim follows immediately.
\end{IEEEproof}

\section{Minimal Networks with Two Source Messages}
\label{sec:twomessages}

In this section we focus on the case of two source messages, i.e.,
$h=2$. Our goal is to show that there exist networks with two source
messages with a positive gap. Such networks have not been demonstrated
in the past, and a gap was shown to exist only in networks with at
least three messages \cite{EtzWac18}. A part of the method we describe
here bears some resemblance to the ones described in
\cite{LehLeh04,FraSol06,ElRGeoSpr06}. The resulting networks are in
fact minimal, as well as sub-networks of the (minimal) combination
network $\cN_{h,r,h}$. Additionally, we show that the networks we
construct attain the largest possible gap of any minimal multicast
network with two source messages.

Of particular interest in what follows, will be the $q$-analog of the
Kneser graph, denoted $qK_{n:m}$, whose set of vertices is
$\sbinom{V}{m}$, where $V=\F_q^n$, and an undirected edge connects two
vertices iff the corresponding $m$-dimensional subspaces have a
trivial intersection (e.g., see \cite{ChoGodRoy06,BloBroSzo12,Ihr19}
and references therein).

We also recall graph homomorphisms, which act as a generalization of
graph coloring. Given two undirected graphs, $\cG_1=(\cV_1,\cE_1)$ and
$\cG_2=(\cV_2,\cE_2)$, we say $\phi:\cV_1\to \cV_2$ is a
\emph{homomorphism}, denoted by abuse of notation $\phi: \cG_1 \to
\cG_2$, if $\mathset{\nu,\nu'}\in \cE_1$ implies
$\mathset{\phi(\nu),\phi(\nu')}\in \cE_2$ for all $\nu,\nu'\in
\cE_1$. This is a generalization of coloring since a homomorphism
$\cG_2\to K_n$ (where $K_n$ denotes the complete graph on $n$
vertices) is equivalent to a coloring of $\cG_2$ with $n$ colors
(think of each of the vertices of $K_n$ representing a distinct color,
which is assigned to the vertices in the reverse image of the
homomorphism). Additionally, it is easy to see that $\chi(\cG_2)$ (the
chromatic number of $\cG_2$) is the minimum $n$ such that there exists
a homomorphism $\cG_2\to K_n$. Since homomorphisms are easily seen to
be transitive, i.e., $\cG_1\to \cG_2 \to \cG_3$ implies $\cG_1\to
\cG_3$, we therefore have that $\cG_1\to \cG_2$ implies
$\chi(\cG_1)\leq \chi(\cG_2)$.

We now continue with the case of two source messages. Let
$\cG=(\cV,\cE)$ be a finite directed acyclic graph. We describe the
following construction of an undirected graph, $\skel(\cG)$, which we
call the \emph{skeleton of $\cG$}.

\begin{construction}
  \label{con:skel}
  Let $\cG=(\cV,\cE)$ be a finite directed acyclic graph. Define
  \begin{align*}
    \cV_{\neq 1} &\eqdef \mathset{ \nu\in\cV ~:~ \din(\nu)\neq 1},\\
    \cE_{\neq 1} &\eqdef \mathset{ (\nu,\nu')\in\cE ~:~ \nu\in \cV_{\neq 1}, \nu'\in\cV}.
  \end{align*}
  Additionally, for each $e\in \cE_{\neq 1}$ we define $T(e)$ to be
  the set of all edges $e'\in\cE$ that may be reached from $e$ by
  directed paths in $\cG$ that pass only through vertices in
  $\cV\setminus \cV_{\neq 1}$. Namely, assuming $e=(\nu_1,\nu_2)$ and
  $e'=(\nu_{m-1},\nu_m)$, then $e'\in T(e)$ if there exist
  $\nu_3,\dots,\nu_{m-2}\in\cV$ such that
  $(\nu_i,\nu_{i+1})\in\cE\setminus\cE_{\neq 1}$ for all $2\leq i\leq
  m-1$.

  We now define the skeleton graph $\skel(\cG)=(\cV',\cE')$. The
  vertex set is defined as
  \[ \cV'=\mathset{ T(e) ~:~ e\in \cE_{\neq 1}}.\]
  The edge set is defined by
  \begin{align*}
    \cE'&=\{\mathset{T(e_1),T(e_2)} ~:~ T(e_1),T(e_2)\in\cV', T(e_1)\neq T(e_2),\\
    &\exists (\nu_1,\nu),(\nu_2,\nu)\in \cE \ \text{s.t.}\  (\nu_1,\nu)\in T(e_1), (\nu_2,\nu)\in T(e_2)\}.
  \end{align*}
\end{construction}

In \cite{LehLeh04} only the reverse process, i.e., mapping what we
call $\skel(\cG)$ back into a network, is used. This is later also
cited and used in \cite{ElRGeoSpr06}. The forward process of mapping a
network to its skeleton bears some resemblance to the procedure
described in \cite{FraSol06}. There, however, a variation on the line
graph is used and not the graph. Additionally, the tree decomposition
described in~\cite{FraSol06} creates a directed graph unlike the
undirected skeleton we have here. The trees in the decomposition
of~\cite{FraSol06} form a subset (sometimes a proper subset) of
$\mathset{T(e) ~:~ e\in \cE_{\neq 1}}$ which was defined here.

A simple observation is proved in the next lemma.

\begin{lemma}
  In Construction~\ref{con:skel}, the set $\cP\eqdef\mathset{T(e) ~:~ e\in
    \cE_{\neq 1}}$ is a partition of $\cE$.
\end{lemma}
\begin{IEEEproof}
  Since $\cG$ is directed and acyclic, find a topological ordering of
  its vertices. Assume to the contrary that $\cP$ is not a partition
  of $\cE$, which implies an edge $e_1=(\nu_1,\nu_2)\in\cE$ is not a
  member of any of the sets in $\cP$ or a member of two sets in
  $\cP$. Assume further, without loss of generality, that among all
  such edges, $e_1$ denotes an edge for which $\nu_1$ is minimal under
  the topological ordering.

  If $\nu_1\in\cV_{\neq 1}$, then by construction, $e_1\in
  T(e_1)\in\cP$ uniquely. Otherwise, there exists exactly a single
  edge $e_0=(\nu_0,\nu_1)\in\cE$. By the minimality of our choice of
  $e_1$, we have $e_0\in T(e)\in\cP$ uniquely, for some $e\in\cE_{\neq
    1}$. By construction, $e_1\in T(e)\in\cP$ uniquely as well.
\end{IEEEproof}

The following lemma connects the existence of a $(q,t)$-linear
solution to the existence of a certain graph homomorphism when we have
$h=2$ source messages.

\begin{lemma}
  \label{lem:skelcolor}
  Let $\cN=(\cG,\sigma,\cT,2)$ be a minimal multicast network. Then
  $\cN$ has a $(q,t)$-linear solution if and only if $\skel(\cG)\to
  qK_{2t:t}$, i.e., there is a homomorphism from $\skel(\cG)$ to
  $qK_{2t:t}$.
\end{lemma}
\begin{IEEEproof}
  In the first direction, assume $\cN$ has a $(q,t)$-linear
  solution. We assume without loss of generality that nodes with
  in-degree $1$ simply forward their incoming message. Thus, given any
  $e\in \cE_{\neq 1}$, all the edges $e'\in T(e)$ carry the same
  message, i.e., $\cM(e')=\cM(e)$. We construct the homomorphism
  $\phi: \skel(\cG)\to qK_{2t:t}$ by setting $\phi(T(e))=\cM(e)$ for
  each $e\in \cE_{\neq 1}$.

  We verify this is indeed a homomorphism, since an edge
  $\mathset{T(e_1),T(e_2)}$ in $\skel(\cG)$ corresponds to edges
  $(\nu_1,\nu)\in T(e_1)$ and $(\nu_2,\nu)\in T(e_2)$ in $\cG$. By
  Corollary~\ref{cor:din}, $\din(\nu)=2$. By Lemma~\ref{lem:dimmv},
  $\dim \cM(\nu)=2t$, hence $\dim \cM(e_1)=\dim \cM(e_2)=t$ and
  $\cM(e_1),\cM(e_2)$ must intersect trivially. Thus, $\cM(e_1)$ and
  $\cM(e_2)$ are joined by an edge in $qK_{2t:t}$.

  In the other direction, assume we have a homomorphism $\phi:
  \skel(\cG)\to qK_{2t:t}$. It is easy to verify that setting
  $\cM(e')=\phi(T(e))$, for all $e'\in T(e)$, $e\in \cE_{\neq 1}$, is
  indeed a $(q,t)$-linear solution, since nodes with in-degree $1$
  simply forward their incoming message, whereas nodes with in-degree
  two receive two trivially intersecting subspaces, and may therefore
  reconstruct the entire space $\F_q^{2t}$. Such nodes can linearly
  create any subspace as an outgoing message.
\end{IEEEproof}

We now prove that the mapping $\skel(\cdot)$ of minimal multicast
networks with two messages to undirected graphs, is surjective. This
already appears in part in \cite{LehLeh04}.

\begin{lemma}
  \label{lem:revskel}
  Let $\cG'=(\cV',\cE')$ be a finite undirected graph with no isolated
  vertices. Then there exists a minimal multicast network
  $\cN=(\cG,\sigma,\cT,2)$ such that $\skel(\cG)=\cG'$.
\end{lemma}
\begin{IEEEproof}
  Construct the desired network $\cN=(\cG,\sigma,\cT,2)$ as
  follows. Let $\cG=(\cV,\cE)$ and fix
  $\cV=\cV'\cup\cE'\cup\mathset{\sigma}$. Additionally, the directed
  edges are defined as
  \[
  \cE = \mathset{(\sigma,\nu) ~:~ \nu\in \cV'} \cup
  \bigcup_{e=\mathset{\nu,\nu'}\in\cE'} \mathset{(\nu,e),(\nu',e)}.
  \]
  The source is $\sigma$, and the terminals are $\cT=\cE'$.

  Intuitively, we construct the network by adding a new source node
  $\sigma$, which is connected to each of the original nodes of
  $\cG'$. Every edge $e=\mathset{\nu,\nu'}$ in the original graph
  $\cG'$ defines a new terminal node into which an edge from $\nu$ and
  an edge from $\nu'$ are connected. It is easy to verify that
  $\skel(\cG)=\cG'$, and that $\cN$ is minimal. The requirement that
  there exist no isolated vertices ensures all nodes in the middle
  layer are essential.
\end{IEEEproof}

We note that the network constructed in Lemma~\ref{lem:revskel} is in
fact a sub-network of the combination network $\cN_{2,r,2}$. Thus, as
a corollary we obtain that the study of minimal multicast networks
with two source messages may be restricted to sub-networks of
combination networks of the form $\cN_{2,r,2}$.

\begin{corollary}
  For any minimal multicast network $\cN=(\cG,\sigma,\cT,2)$ there
  exists a minimal multicast network $\cN'=(\cG',\sigma',\cT',2)$ such
  that $\cG'$ is a subgraph of a combination network, and
  $\gap(\cN)=\gap(\cN')$.
\end{corollary}
\begin{IEEEproof}
  We first note that, by construction, $\skel(\cG)$ does not have isolated
  nodes as these would imply non-essential nodes in $\cG$. We then
  use Lemma~\ref{lem:revskel} to construct $\cN'$ such that
  $\skel(\cG')=\skel(\cG)$. By Lemma \ref{lem:skelcolor}, the networks
  $\cN$ and $\cN'$ have $(q,t)$-linear solutions for exactly the same
  pairs $(q,t)\in\bP\times\N$. Thus, $\gap(\cN)=\gap(\cN')$. Finally,
  as observed, $\cN'$ is a sub-network of a minimal combination
  network $\cN_{2,r,2}$.
\end{IEEEproof}

\begin{figure}
  \begin{center}
    \begin{overpic}[scale=0.45]
      {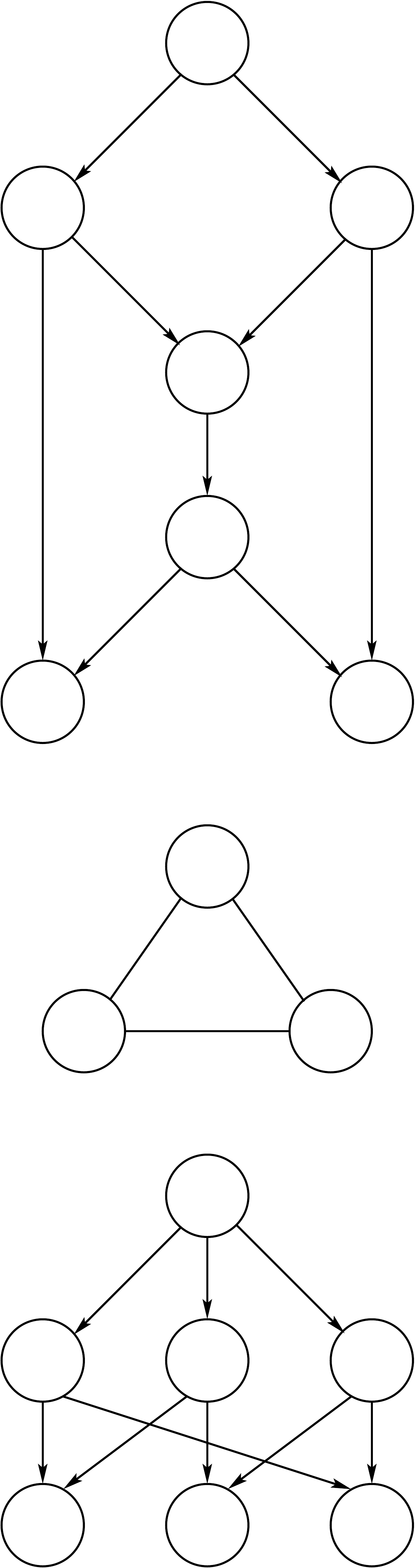}
      \put(12,50){(a)}
      \put(12,30){(b)}
      \put(12,-2){(c)}
      \put(12.7,96.7){$\sigma$}
      \put(2,55){$\tau_1$}
      \put(23,55){$\tau_2$}
      \put(2,86.5){$\nu_1$}
      \put(23,86.5){$\nu_2$}
      \put(12.7,76){$\nu_3$}
      \put(12.7,65.5){$\nu_4$}
      \put(6,94){$e_1$}
      \put(18,94){$e_2$}
      \put(8,83){$e_3$}
      \put(16,83){$e_4$}
      \put(0,71){$e_5$}
      \put(10.5,71){$e_6$}
      \put(21,71){$e_7$}
      \put(6,62){$e_8$}
      \put(18,62){$e_9$}
      \put(3.5,33.75){\small $T(e_1)$}
      \put(11.5,44.25){\small $T(e_6)$}
      \put(19.5,33.75){\small $T(e_2)$}
      \put(7,41){$e'_1$}
      \put(18,41){$e'_2$}
      \put(12,36){$e'_3$}
      \put(12.7,23.2){$\sigma$}
      \put(1,12.7){\small $T(e_1)$}
      \put(11.5,12.7){\small $T(e_6)$}
      \put(22,12.7){\small $T(e_2)$}
      \put(2,2.25){$e'_1$}
      \put(12.5,2.25){$e'_2$}
      \put(23,2.25){$e'_3$}
    \end{overpic}
  \end{center}
  \caption{(a) The famous butterfly network
    $\cN=(\cG,\sigma,\mathset{\tau_1,\tau_2},2)$, (b) its skeleton
    $\skel(\cG)$, and (c) the sub-network of a combination network
    whose skeleton is also $\skel(\cG)$ that is constructed in
    Lemma~\ref{lem:revskel}.}
  \label{fig:skelex}
\end{figure}

\begin{example}
  Figure~\ref{fig:skelex}(a) shows the famous butterfly network,
  $\cN=(\cG,\sigma,\mathset{\tau_1,\tau_2},2)$. For the graph $\cG$ we
  have (in terms of Construction~\ref{con:skel}) that
  \[ \cV_{\neq 1} = \mathset{\sigma,\nu_3,\tau_1,\tau_2}.\]
  Note that we must always have $\sigma\in\cV_{\neq 1}$. Additionally,
  \[ \cE_{\neq 1}=\mathset{e_1,e_2,e_6},\]
  i.e., the set of edges whose source has in-degree that is not
  $1$. We also have
  \begin{align*}
    T(e_1) &= \mathset{e_1,e_3,e_5},\\
    T(e_2) &= \mathset{e_2,e_4,e_7},\\
    T(e_6) &= \mathset{e_6,e_8,e_9}.
  \end{align*}
  Each of these sets becomes a vertex in $\skel(\cG)$, shown in
  Figure~\ref{fig:skelex}(b). Finally, Lemma~\ref{lem:revskel}
  constructs a sub-network of a combination network whose skeleton is
  also $\skel(\cG)$. This is done by adding a source node $\sigma$,
  each vertex in $\skel(\cG)$ becomes a node in the middle layer of
  the constructed graph, and each edge in $\skel(\cG)$ becomes a
  terminal node. The result is depicted in Figure~\ref{fig:skelex}(c).
\end{example}

We can now restate the definition of the gap for minimal multicast
networks $\cN=(\cG,\sigma,\cT,2)$. First, we define
$\cH=\skel(\cG)$. Then, by Lemma~\ref{lem:skelcolor},
\begin{align*}
  q_v(\cN) &= \min \mathset{ q^t ~:~ q\in\bP, t\in\N, \exists \cH\to qK_{2t:t} },\\
  q_s(\cN) &= \min \mathset{ q ~:~ q\in\bP, \exists\cH\to qK_{2:1}}.
\end{align*}
However, since $qK_{2:1}\cong K_{q+1}$, i.e., the complete graph on
$q+1$ vertices, we have
\[ q_s(\cN) = \psi(\chi(\cH)-1),\]
where we use $\psi$ from \eqref{eq:psidef}, since $q_s(\cN)$ is
required to be in $\bP$.

We recall some known results on the chromatic number of certain
$q$-Kneser graphs.

\begin{theorem}
  \label{th:qk1}
  (\cite{ChoGodRoy06,BloBroSzo12,Ihr19})
  Let $q\in\bP$ and $t\in\N$. Then,
  \[\chi(qK_{2t:t}) \leq q^t+q^{t-1}.\]
  If $q\geq 5$ or $t\leq 3$, then
  \[\chi(qK_{2t:t}) = q^t+q^{t-1}.\]
\end{theorem}

We also prove the following lemma which shall become useful soon.

\begin{lemma}
  \label{lem:qkclique}
  Let $q,\oq\in\bP$ and $t,\ot\in\N$. If $q^t> \oq^{\ot}$ then
  \[ qK_{2t:t} \not\to \oq K_{2\ot:\ot}.\]
\end{lemma}
\begin{IEEEproof}
  An upper bound on the size of any clique in $qK_{2t:t}$ is given by
  $(q^{2t}-1)/(q^t-1)=q^t+1$, describing a scenario where (except for
  the zero vector) the subspaces corresponding to the nodes in the
  clique tile $\F_q^{2t}$. This scenario is indeed always possible,
  and is attained by a $t$-spread of $\F_q^{2t}$ (e.g., see
  \cite{SchEtz02}).

  Assume to the contrary that there exists a homomorphism $qK_{2t:t}
  \to \oq K_{2\ot:\ot}$. Since $q$-Kneser graphs never contain self
  loops, the nodes in a clique in $qK_{2t:t}$ are mapped to distinct
  nodes of a clique in $\oq K_{2\ot:\ot}$. But by assumption, $q^t+1>
  \oq^{\ot}+1$, namely, the size of largest clique in $qK_{2t:t}$ is
  strictly larger than the size of the largest clique in $\oq
  K_{2\ot:\ot}$, a contradiction.
\end{IEEEproof}

We are now in a position to provide minimal multicast networks with
two messages with a positive gap. The key to proving this claim is to
build networks whose skeleton is a $q$-Kneser graph. We define these
networks in more generality than required here, since they will be
used in the following section as well.

\begin{definition}
  Let $q\in\bP$, and $h,t\in\N$, $h\geq 2$. We define the \emph{Kneser
    network}, $\cK_{q,t;h}$ in the following way: the network has a
  single source node $\sigma$. Denote $V\eqdef\F_q^{ht}$. The source
  node is connected to $\sbinom{ht}{t}$ nodes in the middle layer,
  which are indexed by
  $\sbinom{V}{t}=\mathset{V_1,V_2,\dots}$. Finally, if we have indices
  $1\leq i_1<i_2<\dots<i_h\leq \sbinom{ht}{t}$ such that
  \begin{equation}
    \label{eq:full}
    V_{i_1}+\dots+V_{i_h} = V = \F_q^{ht},
  \end{equation}
  then their corresponding nodes in the middle layer are connected to
  a unique terminal node.
\end{definition}

We observe that for $h=2$,
\begin{equation}
  \label{eq:kskel2}
  \skel(\cK_{q,t;2}) = qK_{2t:t}.
\end{equation}
We can therefore determine its $q_v$ exactly, which we do in the
following lemma.

\begin{lemma}
  \label{lem:kneserqv}
  For all $q\in\bP$, $t\in\N$,
  \[ q_v(\cK_{q,t;2})=q^t.\]
\end{lemma}
\begin{IEEEproof}
  By \eqref{eq:kskel2}, and since the identity homomorphism
  $qK_{2t:t}\to qK_{2t:t}$ always exists, we have
  \[ q_v(\cK_{q,t;2}) \leq q^t.\]
  By Lemma~\ref{lem:qkclique}, this becomes an equality, as claimed.
\end{IEEEproof}

We now show a gap exists in many Kneser networks with two source
messages.

\begin{theorem}
  \label{th:gap1}
  For all $q\in\bP$ and $t\in\N$, with $q\geq 5$ or $t\leq 3$,
  \[\gap(\cK_{q,t;2}) = \psi(q^t+q^{t-1}-1)-q^t \geq q^{t-1}-1.\]
\end{theorem}
\begin{IEEEproof}
  By Lemma~\ref{lem:kneserqv},
  \[ q_v(\cK_{q,t;2}) = q^t.\]
  However, by Theorem~\ref{th:qk1}, the chromatic number of
  $qK_{2t:t}$ is
  \[ \chi(qK_{2t:t})=q^t+q^{t-1},\]
  for $q\geq 5$ or $t\leq 3$. Thus, for these cases
  \[ q_s(\cK_{q,t;2}) = \psi(q^t+q^{t-1}-1),\]
  and the equality part of the claim is proved. The claimed inequality
  follows from $\psi(n)\geq n$ for all $n\in\N$.
\end{IEEEproof}

This result is matched by the following upper bound.

\begin{theorem}
  \label{th:gap1up}
  If $\cN=(\cG,\sigma,\cT,2)$ is a minimal multicast network with a
  $(q,t)$-vector-optimal linear solution, then
  \[\gap(\cN)\leq \psi(q^t+q^{t-1}-1)-q^t\leq q^t+2q^{t-1}-2.\]
\end{theorem}
\begin{IEEEproof}
  By definition, a $(q,t)$-vector-optimal linear solution implies
  $q_v(\cN)=q^t$, and we have
  \[ \skel(\cG) \to qK_{2t:t}.\]
  By Theorem~\ref{th:qk1} we know that $\chi(qK_{2t:t})\leq q^t+q^{t-1}$.
  We denote $q'=\psi(q^t+q^{t-1}-1)$, and therefore
  \[ qK_{2t:t}\to K_{\chi(qK_{2t:t})}\to K_{q^t+q^{t-1}} \to K_{q'+1}\cong q'K_{2:1},\]
  where the third homomorphism follows trivially from the fact that
  $q'+1\geq q^t+q^{t-1}$. By the transitivity of homomorphisms we have
  \[ \skel(\cG) \to q'K_{2:1},\]
  and then by Lemma~\ref{lem:skelcolor}, the network $\cN$ has a
  $(q',1)$-linear solution, namely,
  \[ q_s(\cN)\leq q' = \psi(q^t+q^{t-1}-1).\]
  Finally, the last inequality follows from \eqref{eq:primelarge}.
\end{IEEEproof}

\begin{corollary}
  Kneser networks with two source messages, $\cK_{q,t;2}$, $q\in\bP$,
  $t\in\N$, and $q\geq 5$ or $t\leq 3$, attain the maximum possible
  gap among all minimal networks with two source messages and
  $(q,t)$-vector-optimal linear solutions.
\end{corollary}

A few cases remain uncovered by our previous treatment, namely,
$q\in\mathset{2,3,4}$ and $t\geq 4$. We can show a gap for these cases
as well, albeit a much smaller guaranteed gap of merely~$1$.

\begin{theorem}
  For all $q\in\bP$ and all $t\in\N$, $t\geq 2$,
  \[\gap(\cK_{q,t;2})\geq \psi(q^t+1)-q^t \geq 1.\]
\end{theorem}
\begin{IEEEproof}
  Again, by Lemma~\ref{lem:kneserqv}, $q_v(\cK_{q,t;2})=q^t$. We now
  contend that $\chi(qK_{2t:t})> q^t+1$. Recall that the vertex set of
  $qK_{2t:t}$ is $\sbinom{V}{t}$, where $V=\F_q^{2t}$. Assume a
  coloring of $qK_{2t:t}$ with $c$ colors. Let $U_i\subset
  \sbinom{V}{t}$, $1\leq i\leq c$, be the set of vertices colored with
  color $i$. Then each $U_i$ is a $1$-intersecting family in the
  language of \cite{FraWil86}\footnote{If $V\eqdef\F_q^n$, then
    $\cF\subseteq\sbinom{V}{\ell}$ is an \emph{$m$-intersecting
      family} if for all $W,W'\in\cF$, $\dim(W\cap W')\geq m$.}, and
  an anticode of diameter $t-1$ in the language of
  \cite{SchEtz02}\footnote{If $V\eqdef\F_q^n$, the distance between
    $W,W'\in\sbinom{V}{\ell}$ is defined as $d(W,W')\eqdef
    \ell-\dim(W\cap W')$. An anticode of diameter $m$ is a set
    $\cF\subseteq\sbinom{V}{\ell}$ such that $W,W'\in\cF$ implies
    $d(W,W')\leq m$.}. Also, the set $\mathset{U_i}_{1\leq i\leq c}$
  forms a tiling (partition) of $\sbinom{V}{t}$.

  By \cite{FraWil86}, for all $1\leq i\leq c$,
  \[ \abs{U_i} \leq \sbinom{2t-1}{t-1},\]
  and $U_i$ is either
  \[ U_i = \mathset{ U\in\sbinom{V}{t} ~:~ V_1\subseteq U },\]
  or
  \[ U_i = \mathset{ U\in\sbinom{V}{t} ~:~ U\subseteq V_{2t-1}},\]
  where $V_1,V_{2t-1}$ are subspaces of $V$ of dimensions $1$ and
  $2t-1$, respectively. However, by \cite{SchEtz02}, there is no
  tiling of $V$ by $U_i$ of these shapes. Thus,
  \[ \chi(qK_{2t:t}) > \frac{\sbinom{2t}{t}}{\sbinom{2t-1}{t-1}} = q^t+1.\]
  It follows that
  \[ \gap(\cK_{q,t;2})=q_s(\cK_{q,t;2})-q_v(\cK_{q,t;2}) \geq \psi(q^t+1)-q^t \geq 1,\]
  as claimed.
\end{IEEEproof}

\section{Minimal Networks with More Than Two Source Messages}
\label{sec:moremessages}

Prior to this work, a gap was shown to exist only in networks with at
least four source messages \cite{EtzWac18}, and the networks there are
not minimal.  An additional ad-hoc example with three source messages
was also given in \cite{EtzWac18} but not in the form of a
gap. Instead, it was shown that for the same field size, a different
number of nodes in the middle layer of the network is required.

In the previous section we showed a gap exists for $h=2$ source
messages.  We first contend this immediately shows a gap exists in
networks with any $h\geq 3$ source messages. We show this by giving a
general construction that translates any network $\cN$ with $h$ source
messages to a network $\cN'$ with $h'>h$ source messages while keeping
the parameters of the solution.

\begin{construction}
  \label{con:extend}
  Let $\cN=(\cG,\sigma,\cT,h)$ be a multicast network, with
  $\cG=(\cV,\cE)$, and $h\in\N$. For any $h'\in\N$, $h'>h$, we
  construct the network $\cN'=(\cG',\sigma',\cT',h')$, with
  $\cG'=(\cV',\cE')$, as follows.

  We set $\cV'=\cV\cup\mathset{\sigma'}$, and $\cT'=\cT$. We keep all
  of the edges of $\cE$ in $\cE'$, and we add $h$ parallel edges from
  $\sigma'$ to $\sigma$. Finally, for each terminal $\tau\in\cT'$, we
  add $h'-h$ parallel edges from $\sigma'$ to $\tau$.
\end{construction}

Intuitively, Construction~\ref{con:extend} keeps the network $\cN$ in
its entirety. The terminals also remain as they were but a new source
is added. The new source is connected to the original source by $h$
parallel edges, and to each of the terminals by $h'-h$ parallel edges.
We also make the observation that if $\cN$ is minimal, then so is
$\cN'$.

\begin{lemma}
  \label{lem:extendkeepsol}
  Let $h,h'\in\N$, $h'>h$. If $\cN=(\cG,\sigma,\cT,h)$ and
  $\cN'=(\cG',\sigma',\cT',h')$ are as in
  Construction~\ref{con:extend}, then there exists a $(q,t)$-linear
  solution to $\cN$ if and only if there exists a $(q,t)$-linear
  solution to $\cN'$.
\end{lemma}
\begin{IEEEproof}
  In the first direction, assume we have a $(q,t)$-linear solution to
  $\cN$. We build a simple $(q,t)$-linear solution to $\cN'$ by using
  the same messages over the edges in $\cE$, sending the $h$ source
  messages $x_1,\dots,x_h$ from $\sigma'$ to $\sigma$ over the new $h$
  edges, and sending $x_{h+1},\dots,x_{h'}$ over the $h'-h$ edges
  connecting $\sigma'$ to each of the terminals.

  In the second direction, assume we have a $(q,t)$-linear solution to
  $\cN'$. By change of bases we can assume without loss of generality
  that $x_1,\dots,x_h$ are sent over the $h$ edges connecting
  $\sigma'$ to $\sigma$. Consider any terminal $\tau\in\cT$. Since
  only $h'-h$ new parallel edges connect $\sigma'$ to $\tau$ directly,
  the remaining incoming edges into $\tau$, namely,
  $\inc(\tau)\cap\cE$, must contain messages that enable the recovery
  of $x_1,\dots,x_h$. Hence, restricting ourselves to the original
  network, there is a $(q,t)$-linear solution allowing $\sigma$ to
  convey all of its $h$ messages to the terminals.
\end{IEEEproof}

\begin{corollary}
  For any network $\cN=(\cG,\sigma,\cT,h)$ there exists a network
  $\cN'=(\cG',\sigma',\cT',h')$ with $h'>h$ such that
  \[ \gap(\cN)=\gap(\cN').\]
\end{corollary}

It now follows that there exist networks with $h>2$ source messages
that exhibit the same gap as the networks we constructed in the
previous section with $h=2$ source messages. Moreover, the minimality
of the networks is preserved. However, the networks resulting from
Construction~\ref{con:extend} are not combination networks, nor are
they sub-networks of combination networks. Under the assumption that
such a structure might be of possible interest, theoretical or
practical, we proceed by generalizing the previous section's results
to more than two source messages.

A full-fledged generalization of the case of $h=2$ to $h>2$ via the
skeleton-graph approach, seems highly intricate. Instead, we study
Kneser networks $\cK_{q,t;h}$ with $h > 2$. Their analysis is made
possible by replacing their skeleton, a $q$-analog of the Kneser
graph, with a generalization of it to a hypergraph.

We assume throughout this section that $h>2$. We again observe that
$\cK_{q,t;h}$ is a minimal multicast network which is a sub-network of
a minimal combination network of the type $\cN_{h,r,h}$. We also
observe that trivially,
\begin{equation}
  \label{eq:trivh3}
  q_v(\cK_{q,t;h})\leq q^t,
\end{equation}
since a simple linear solution is for the source $\sigma$ to send the
vector space $V_i\in\sbinom{V}{t}$ to the node in the middle layer
indexed by $V_i$, and all the nodes in the middle layer just forward
their incoming message.

It remains to consider a scalar solution for $\cK_{q,t;h}$. To that
end we define the $q$-analog Kneser \emph{hypergraph}. The $q$-analog
Kneser hypergraph, denoted $qK^h_{ht:t}$ has
$\sbinom{V}{t}=\mathset{V_1,V_2,\dots}$ as vertices, and an
$h$-hyperedge $\mathset{V_{i_1},V_{i_2},\dots,V_{i_h}}$ exists, if and
only if \eqref{eq:full} holds. We note that our generalization to the
$q$-analog Kneser hypergraph is different from other generalizations,
e.g.,~\cite{Meu14} and the references therein.

Several definitions exist for colorings of hypergraphs. In our case, a
coloring is an assignment of a color to each of the vertices of the
hypergraph such that no hyperedge contains two vertices of the same
color. The minimal number of colors required to color a given
hypergraph $\cG$ is called its \emph{chromatic number}, and is denoted
by $\chi(\cG)$.

\begin{lemma}
\label{lem:h_color}
For all $q\in\bP$, $t,h\in\N$, and $h \geq 3$,
\[\chi(qK^h_{ht:t})=\chi(qK_{ht:t}).\]
\end{lemma}
\begin{IEEEproof}
  We prove every coloring of $qK^h_{ht:t}$ is a coloring of
  $qK_{ht:t}$, and vice versa. First, consider a coloring $c$ of
  $qK_{ht:t}$. We contend it is also a coloring of
  $qK^h_{ht:t}$. Indeed, for any $h$ subspaces of dimension $t$,
  $V_{i_1},\dots,V_{i_h}\subseteq V=\F_q^{ht}$, such that
  \eqref{eq:full} holds, each pair of them is trivially intersecting,
  hence $c$ gives them all different colors.

  In the other direction, assume $c$ is a coloring of $qK^h_{ht:t}$
  and we prove it is a coloring of $qK_{ht:t}$. Given two trivially
  intersecting subspaces of dimension $t$, $V_{i_1},V_{i_2}$, we can
  easily build $h-2$ more subspaces of dimension $t$,
  $V_{i_3},\dots,V_{i_h}$, such that \eqref{eq:full} holds. Thus,
  $V_{i_1}$ and $V_{i_2}$ are elements in an hyperedge of
  $qK^h_{ht:t}$, hence $c$ colors them distinctly.
\end{IEEEproof}

We recall more results on the chromatic number of $qK_{n:m}$.
\begin{theorem} \cite{ChoGodRoy06,BloBroChoFraPatMusSzo10}
  \label{th:qk2}
  Let $q\in\bP$, and $n,m\in\N$, $n\geq 2m+1$. Except for the
  case of $n=2m+1$ and $q=2$, in all other cases,
  \[ \chi(qK_{n:m}) = \sbinom{n-m+1}{1} = \frac{q^{n-m+1}-1}{q-1}.\]
\end{theorem}

We can now state the existence of a gap for Kneser networks with more
than two source messages.

\begin{theorem}
  For all $q\in\bP$, $t,h\in\N$, $t\geq 2$, and $h\geq 3$,
  \begin{align*}
    \gap(\cK_{q,t;h})
    &\geq \begin{cases}
      \psi\parenv{q^t+\frac{1}{h-1}q^{t-1}}-q^t & t\geq h, \\
      \psi\parenv{q^t+\frac{1}{(h-1)^2}q^{t-1}}-q^t & \text{otherwise.}
    \end{cases}\\
    &\geq \begin{cases}
      \frac{1}{h-1}q^{t-1} & t\geq h, \\
      \frac{1}{(h-1)^2}q^{t-1} & \text{otherwise.}
    \end{cases}
  \end{align*}
\end{theorem}
\begin{IEEEproof}
  As already observed in \eqref{eq:trivh3},
  \[ q_v(\cK_{q,t;h})\leq q^t.\]
  For the scalar solution, our choice of $\cK_{q,t;h}$ allows us to
  follow in a manner similar to the previous section. To avoid tedious
  notation, denote $q_s\eqdef q_s(\cK_{q,t;h})$. The source $\sigma$
  sends a one-dimensional subspace of $\F_{q_s}^h$ on each of the
  links to the nodes in the middle layer. We may think of the choice
  of subspace as a color which we assign to the nodes in the middle
  layer. The total number of colors at our disposal is
  \[\sbinom{h}{1}_{q_s} = \frac{q_s^h-1}{q_s-1}.\]
  The structure of $\cK_{q,t;h}$ implies that a scalar solution must
  induce a valid coloring of $qK^h_{ht:t}$. Therefore, using
  Lemma~\ref{lem:h_color} and Theorem~\ref{th:qk2},
  \[ \frac{q_s^h-1}{q_s-1} \geq \chi(qK^h_{ht:t})=\chi(qK_{ht:t}) = \frac{q^{(h-1)t+1}-1}{q-1}.\]
  If we define
  \[ f(x)\eqdef \frac{x^h-1}{x-1}=\sum_{i=0}^{h-1} x^i,\]
  then, since $f(x)$ is strictly increasing,
  \[ q_s \geq \sup\mathset{ x\in\R ~:~ f(x) \leq \frac{q^{(h-1)t+1}-1}{q-1}}.\]

  If $t\geq h$ we observe that for all $1\leq i\leq h-1$,
  \begin{align*}
    \parenv{q^t+\frac{1}{h-1}\cdot q^{t-1}}^i 
    & = \sum_{j=0}^i \binom{i}{j}\frac{1}{(h-1)^j}\cdot q^{ti-j}\\
    & \leq \sum_{j=0}^i q^{ti-j}
  \end{align*}
  which follows from
  \[\binom{i}{j}\frac{1}{(h-1)^j}\leq \parenv{\frac{i}{h-1}}^j\leq 1,\]
  and $i\leq h-1$. Thus,
  \begin{align*}
    f\parenv{q^t+\frac{1}{h-1}\cdot q^{t-1}}
    &= \sum_{i=0}^{h-1} \parenv{q^t+\frac{1}{h-1}\cdot q^{t-1}}^i\\
    &\leq \sum_{i=0}^{h-1}\sum_{j=0}^i q^{ti-j} \\
    &\leq \sum_{i=0}^{(h-1)t}q^i \\
    &= \frac{q^{(h-1)t+1}-1}{q-1},
  \end{align*}
  where the last inequality uses the fact that $t\geq h$ and hence no
  power of $q$ repeats in the last double summation. It follows that
  when $t\geq h$ we have
  \[q_s \geq \psi\parenv{q^t+\frac{1}{h-1}\cdot q^{t-1}}.\]

  The case of $t<h$ requires a slightly different treatment. For this
  case, when $1\leq i\leq h-1$,
  \begin{align*}
    \parenv{q^t+\frac{1}{(h-1)^2}\cdot q^{t-1}}^i 
    & = \sum_{j=0}^i \binom{i}{j}\frac{1}{(h-1)^{2j}}\cdot q^{ti-j}\\
    &\overset{(a)}{\leq} q^{ti} + \frac{i}{h-1}\cdot q^{ti-1}\\
    &\overset{(b)}{\leq} q^{ti} + q^{ti-1},
  \end{align*}
  where (a) follows from
  \[\binom{i}{j}\frac{1}{(h-1)^{2j}}\leq \parenv{\frac{i}{(h-1)^2}}^j\leq \frac{1}{h-1},\]
  since $i\leq h-1$, and (b) follows again from $i\leq h-1$. Thus,
  \begin{align*}
    f\parenv{q^t+\frac{1}{(h-1)^2}\cdot q^{t-1}}
    &= \sum_{i=0}^{h-1} \parenv{q^t+\frac{1}{(h-1)^2}\cdot q^{t-1}}^i\\
    &\leq 1+\sum_{i=1}^{h-1}(q^{ti}+q^{ti-1}) \\
    &\leq \sum_{i=0}^{(h-1)t}q^i \\
    &= \frac{q^{(h-1)t+1}-1}{q-1}.
  \end{align*}
  It follows that when $t<h$ we have
  \[q_s \geq \psi\parenv{q^t+\frac{1}{(h-1)^2}\cdot q^{t-1}}.\]
  By combining the bounds on $q_v$ and $q_s$ in all the cases we
  obtain the desired result.
\end{IEEEproof}

\section{Minimal Full Combination Networks}
\label{sec:mincomb}

We turn to consider, in this section, an upper bound on the gap. In
particular, we study the full minimal combination network
$\cN_{h,r,h}$, and show that it has a very small gap (if at all)
compared to its sub-networks, which were considered in the previous
section.

The key result we use is the following theorem proved
in~\cite{RiiAhl06}.  Let $(r,q^h,r-s+1)_q$ denote a code over $\F_q$
of length $r$ with $q^h$ codewords and minimum Hamming distance
$r-s+1$. If this code is linear, it is denoted by $[r,h,r-s+1]_q$.

\begin{theorem} (\cite{RiiAhl06})
\label{thm:RiiAhl}
The $\mathcal{N}_{h,r,s}$ combination network is solvable over $\F_q$
if and only if there exists an $(r,q^h,r-s+1)_q$ code.
\end{theorem}

In view of Theorem~\ref{thm:RiiAhl}, what are the functions on the edges
of the $\mathcal{N}_{h,r,s}$ combination network in the three types of
solutions -- nonlinear, vector linear, and linear?

For the (scalar) \emph{nonlinear} solution, take an $(r,q^h,r-s+1)_q$
code. We can feed the $h$ source messages to an arbitrary encoder for
the code to obtain a codeword. The $r$ symbols of the codeword are
then transmitted on the $r$ links leaving the source node. In the
middle layer, each node simply repeats its incoming message on all of
its outgoing links. Finally, each terminal obtains $s$ symbols from
the codeword, and since the code has minimum distance of $r-s+1$ it
may recover the entire codeword from the surviving $s$ symbols, and
reverse the encoding process to obtain the $h$ source messages.

For the scalar \emph{linear} solution, an $[r,h,r-s+1]_q$ code is
required. We use the same approach as the one for nonlinear solutions,
but using a linear encoder for the code results in a scalar linear
solution. Namely, the code has an $r \times h$ generator matrix and
the $h$~entries of its $i$th row are the coding coefficients of the
linear function on the link from the source to the $i$th node in the
middle layer.

The vector-linear case differs. A fundamental combinatorial structure
that underpins the vector solutions to minimal combination networks is
a structure we call a \emph{$(t;h,\alpha)_q$-independent
  configuration}.

\begin{definition}
Let $q\in\bP$, $t,h,\alpha\in\N$, $\alpha \leq h$, and denote
$V=\F_q^{ht}$. A \emph{$(t;h,\alpha)_q$-independent configuration
  (IC)} is a set
$\cC=\mathset{V_1,V_2,\dots,V_m}\subseteq\sbinom{V}{t}$, such that for
all $1\leq i_1 < i_2 < \dots < i_{\alpha}\leq m$,
\[ \dim(V_{i_1}+V_{i_2}+\dots+V_{i_\alpha})=\alpha t.\]
We say $\abs{\cC}=m$ is the \emph{size} of the IC.
\end{definition}

We now make the connection between ICs and full minimal combination
networks, $\cN_{h,r,h}$.

\begin{lemma}
\label{lem:icsol}
The $\cN_{h,r,h}$ combination network has a $(q,t)$-solution over
$\F_q$ if and only if there exists a $(t;h,h)_q$-IC of size $r$.
\end{lemma}
\begin{IEEEproof}
In the first direction assume that a $(q,t)$-solution over $\F_q$
exists. We note that by construction, any node $i$ in the middle layer
of $\cN_{h,r,h}$ receives a subspace $V_i\subseteq V\eqdef\F_q^{ht}$,
with $\dim(V_i) \leq t$.  If the terminal $R_j$ gets from the middle
layer the subspaces $V_{i_1},\dots,V_{i_{h}}$, then
\[
\dim(V_{i_1}+\dots+V_{i_h}) =ht,
\]
which implies that $\dim(V_i) = t$. Thus, $\mathset{V_i}_{1\leq i\leq
  r}$ is a $(t;h,h)_q$-IC.

In the other direction, assume $\cC=\mathset{V_1,\dots,V_r}$ is a
$(t;h,h)_q$-IC. We can easily construct a vector network coding
solution to the $\cN_{h,r,h}$ combination network. Simply send $V_i$
to the $i$th middle layer node. Since $\cC$ is a $(t;h,h)_q$-IC it
follows that each terminal has a full rank $(ht) \times (ht)$ transfer
matrix from which it can recover the $h$ messages.
\end{IEEEproof}

We now bound the size of ICs, which will later enable us to upper
bound the gap in minimal combination networks.

\begin{lemma}
\label{lem:icsize}
Let $\cC$ be a $(t;h,\alpha)_q$-IC. If $\alpha\geq 2$ then
\[
\abs{\cC} \leq \frac{q^{(h-\alpha+2)t}-1}{q^{t}-1}+\alpha-2.
\]
\end{lemma}
\begin{IEEEproof}
If $\alpha=2$ the claim is immediate by considering the size of a
$t$-spread (e.g., see \cite{SchEtz02}).

Assume now $\alpha>2$, and denote $V\eqdef \F_q^{ht}$. Let us write
$\cC=\mathset{V_1,V_2,\dots,V_m}$, and define
\[W_1\eqdef V_1 + V_2 +\dots +V_{\alpha-2},\]
where $\dim(W_1)=(\alpha-2) t$.  By the definition of an IC,
$\F_q^{ht}=W_1 + W_2$, where $W_2\in\sbinom{V}{(h-\alpha+2)t}$. It
follows that any vector $v\in V_j$, $\alpha-1\leq j\leq m$, may be
written uniquely as $v=v_1+v_2,$ where $v_1\in W_1$ and $v_2\in
W_2$. We now define
\[V'_j\eqdef\mathset{v_2 ~:~ v_1+v_2\in V_j,\, v_1\in W_1,\, v_2\in W_2},\]
for all $\alpha-1\leq j\leq m$. It is easily seen that
$\dim(V'_j)= t$.

Furthermore, for any $\alpha-1\leq j_1<j_2\leq m$,
\[
\dim(W_1+V'_{j_1}+ V'_{j_2})=\alpha t \Rightarrow \dim(V'_{j_1}+V'_{j_2})=2 t.
\]
Thus, the set $\mathset{V'_i}_{\alpha-1\leq i\leq m}$ contains $\abs{\cC}-\alpha+2$ pairwise disjoint $t$-subspaces of $W_2$.
Thus,
\[
\abs{\cC}-\alpha+2 \leq \frac{\sbinom{(h-\alpha+2)t}{1}}{\sbinom{ t}{1}}.
\]
\end{IEEEproof}

When $t=1$, bounding the size of $(1;h,h)_q$-IC is equivalent to
finding the longest MDS codes, and hence related to the MDS
conjecture. Thus, Lemma~\ref{lem:icsize} forms a generalization for an
upper bound on the length of MDS code. The related results for
(scalar, $t=1$) linear codes are given
in~\cite{MacSlo78}. Corollary~7~\cite[p. 321]{MacSlo78} asserts that
for an $[n,k,n-k+1]_q$ MDS code, we have that $n \leq q + k -1$. This
result is strengthened in Theorem~11~\cite[p. 326]{MacSlo78} by using
a more complicated proof based on projective geometry. The theorem
asserts that if $k \geq 3$ and $q$ is odd then $n \leq q+k-2$.
Another involved proof for the same result is given for nonlinear
codes in~\cite[pp. 12-13]{Rag71}.

We can now give an upper bound on the gap for minimal combination
networks.

\begin{theorem}
  \label{th:gapcomb}
  For all $h,r\in\N$, $r\geq h\geq 2$,
  \[\gap (\cN_{h,r,h})\leq \psi(r-1)-\psi(r-h+1) \leq r+h-3,\]
  and for all large enough $r$,
  \[\gap(\cN_{h,r,h}) \leq (r-1)^{21/40}+h-2.\]
\end{theorem}

\begin{IEEEproof}
  By \cite{RiiAhl06}, a $(q,1)$-scalar linear solution to $\cN_{h,r,h}$ exists
  if and only if an $[r,h,r-h+1]_{q_s}$ MDS code exists. Thus, we
  certainly have
  \[q_s(\cN_{h,r,h})\leq \psi(r-1),\]
  (e.g., see \cite[pp.~317--331]{MacSlo78}). On the other hand, by
  Lemma~\ref{lem:icsol}, and Lemma~\ref{lem:icsize}
  \[ q_v(\cN_{h,r,h})\geq \psi(r-h+1).\]
  Thus, by \eqref{eq:primelarge}, for all $r\in\N$,
  \begin{align*}
    \gap(\cN_{h,r,h}) &\leq \psi(r-1)-\psi(r-h+1)\\
    &= (\psi(r-1)-(r-1)) \\
    &\quad\ -(\psi(r-h+1)-(r-h+1))+h-2 \\
    &\leq r+h-3,
  \end{align*}
  and by \eqref{eq:primesmall}, for all large enough $r$,
  \[ \gap(\cN_{h,r,h}) \leq (r-1)^{21/40}+h-2.\]
\end{IEEEproof}

Loosely speaking, for the minimal combination network, $\cN_{h,r,h}$,
the gap is upper bounded by $h-2$ plus the maximum distance between
powers of primes. Since it is conjectured that $\psi(n)-n\leq O(\log
n)$, the upper bound may be further reduced in the future if this
conjecture is proved.

\section{Conclusion}
\label{sec:conclude}

In this paper we studied scalar and vector linear solutions to the
combination network and its sub-networks. We first found the maximal
possible gap for minimal multicast networks with $h=2$ source
messages, and showed that the Kneser network, which was defined in this
work, attains this bound on the gap with equality. Furthermore, we
studied Kneser networks with $h\geq 3$ source messages and proved they
also exhibit a gap. Finally, we provided an upper bound on the gap of
full minimal combination networks, showing the gap is relatively
small.

It is interesting to compare the gap results with those of
\cite{EtzWac18}. In~\cite{EtzWac18} it was proved that for even $h\geq
4$ there exists a multicast network $\cN$ for which $\gap(\cN) =
q^{(h-2)t^2/h +o(t)}$, and for odd $h\geq 5$ there exists a multicast
network $\cN$ for which $\gap(\cN) = q^{(h-3)t^2/(h-1) +o(t)}$, and in
any case, the gap is of order $q^{\Theta(t^2)}$, where $q$ and $h$ are
considered constant. In comparison with the results of Section
\ref{sec:twomessages} and Section~\ref{sec:moremessages}, the gap we
present is only of the order of $q^{\Theta(t)}$, which we also prove
is optimal in the case of minimal multicast networks with $h=2$. Two
important points distinguish between the two papers: the networks in
\cite{EtzWac18} are not minimal, and \cite{EtzWac18} uses $h\geq
4$. Which of the two, minimality or the number of source messages,
contributes to this difference is still an open question. The results
of Section \ref{sec:moremessages} may hint at the former, but we are
still missing an upper bound for $h\geq 3$.

We also want to highlight the significance of Section
\ref{sec:mincomb}. If we take for example the case of $h=2$ source
messages, by Theorem~\ref{th:gapcomb}, the full minimal combination
network $\cN_{2,r,2}$ has~$0$ gap, for any $r\in\N$ (a fact that was
noticed in \cite{EtzWac18} using Latin squares). If we pick $q\in\bP$,
$t\in\N$, $r=\sbinom{2t}{t}$, keep the source and all of the middle
layer of nodes, but remove a few of the terminals, we obtain the
Kneser network $\cK_{q,t;2}$, the network from Theorem~\ref{th:gap1},
that has the maximal possible gap, $\psi(q^t+q^{t-1}-1)-q^t$. Thus, it
appears that $\cN_{2,r,2}$ is equally difficult for a scalar solution
as it is for a vector solution. However, by pruning some terminals,
the resulting Kneser network $\cK_{q,t;2}$ becomes easy for a vector
solution but difficult for a scalar solution. This upper bound on the
gap by Theorem~\ref{th:gapcomb} grows very slowly. As another example,
if $r-2$ is not a prime power then $\cN_{3,r,3}$ has no gap.

A full characterization of the gap in linear multicast networks is
still an open question. We also suggest minimality of such networks
may play a role in limiting the gap, a problem which we leave for
future work. Finally, the gap between non-linear and linear solutions
for multicast networks is still unknown.

\section*{Acknowledgment}
The authors thank Yiwei Zhang for helpful discussions.

\bibliography{allbib}
\bibliographystyle{IEEEtranS}

\end{document}

%% file: comb-network.tex
\centering
\tikzsetnextfilename{2-butterflies}
\begin{tikzpicture}[scale = 1]
	\node[mycircle,label=right:{$x_1,\dots,x_h$}] (sourcex) {} ; 
	\node[mycircle,below left=30pt and 40pt of sourcex] (middle0) {} ;
	\node[mycircle,below left=30pt and 20pt of sourcex] (middle1) {} ;
	\node[mycircle,below left=30pt and 0pt of sourcex] (middle2) {} ;
	\node[rectangle,right = 10pt of middle2](text1){$\dots$};
	\node[mycircle,below left=30pt and -60pt of sourcex] (middle3) {} ;
	\node[mycircle,below left=30pt and -80pt of sourcex] (middle4) {} ;
	\node[rectangle,right = 5pt of middle4](text2){$r$ nodes};

	\draw[black,->,very thick] (sourcex.south) -- (middle0.north);
	\draw[black,->,very thick] (sourcex.south) -- (middle1.north);
	\draw[black,->,very thick] (sourcex.south) -- (middle2.north);
	\draw[black,->,very thick] (sourcex.south) -- (middle3.north);
	\draw[black,->,very thick] (sourcex.south) -- (middle4.north);

	\draw [-, very thick,black] (2.2,-2.3) arc (160:50:10pt);
	\draw [-, very thick,black] (-1.8,-2.4) arc (140:30:13pt);
	\draw [-, very thick,black] (-2.5,-2.3) arc (150:20:10pt);
	
	\node[mycircle,below left=30pt and 10pt of middle0] (rec1) {} ;
	\node[mycircle,below left=30pt and -20pt of middle0] (rec2) {} ;
	\node[mycircle,below left=30pt and -150pt of middle0] (reclast) {} ; 
	
	\draw[black,->,very thick] (middle0.south) -- (rec1.north);
	\draw[black,->,very thick] (middle1.south) -- (rec1.north);
	\draw[black,->,very thick] (middle0.south) -- (rec2.north);
	\draw[black,->,very thick] (middle2.south) -- (rec2.north);
	\draw[black,->,very thick] (middle3.south) -- (reclast.north);
	\draw[black,->,very thick] (middle4.south) -- (reclast.north);
	\node[rectangle,below right = 7pt and 7pt of middle4](text2){$s$ edges};
	\node[rectangle,below right = 28pt and 20pt of middle4](text2){$\binom{r}{s}$ receivers};
	
	\node[rectangle,right = 23pt of middle2](blub){$\dots$};
		\draw[black,->,very thick] (blub.south) -- (reclast.north);
	
\end{tikzpicture}